\documentclass[doublespacing]{elsart}
\usepackage{graphicx}
\newcommand{\Vec}[1]{\mbox{\boldmath$#1$}}

\begin{document}

\begin{frontmatter}

\title{Unconventional pairing originating from 
disconnected Fermi surfaces in the iron-based superconductor}

\author[address1,address3]{Kazuhiko Kuroki\corauthref{Kuroki}},
\ead{kuroki@vivace.e-one.uec.ac.jp}
\author[address2,address3]{Hideo Aoki}

\address[address1]{Department of Applied Physics and Chemistry, 
The University of Electro-Communications, 
Chofu, Tokyo 182-8585, Japan}

\address[address2]{Department of Physics,  
University of Tokyo, Hongo, Tokyo 113-0033, Japan}

\address[address3]{JST, TRIP, Sanbancho, Chiyoda, Tokyo 102-0075, Japan}

\corauth[Kuroki] {Corresponding author. Tel: +81 (42) 486-9036}

\begin{abstract}
For the iron-based high $T_c$ superconductor 
LaFeAsO$_{1-x}$F$_x$, we construct a minimal model, where 
all of the five Fe $d$ bands turn out to be involved.  
We then investigate the origin of superconductivity 
with a five-band random-phase approximation by solving 
the Eliashberg equation.  We conclude that 
the spin fluctuation modes arising from 
the nesting between the disconnected Fermi 
pockets realise, basically, an extended $s$-wave pairing, 
where the gap changes sign across the nesting vector.
\end{abstract}

\begin{keyword}
Iron pnictide superconductors, disconnected Fermi surface, 
spin fluctuation mediated pairing
\PACS 74.20.Mn
\end{keyword}
\end{frontmatter}

\section{Introduction}
While the physics of high-$T_c$ cuprate has matured after 
the two decades since the discovery, 
the superconductivity 
the iron-based pnictide LaFeAsO doped with fluorine 
discovered by Hosono's group\cite{Hosono} 
is more than welcome as a fresh challenge for yet 
another class of high-$T_c$ systems.  Indeed,  
the iron-based material, 
along with various other ones in the same family of compounds 
with higher transition temperatures ($T_c$)\cite{Ren}, 
are remarkable as the first non-copper compound that 
has $T_c$'s exceeding 50 K.  This immediately 
stimulates renewed interests in the 
electronic mechanism of high $T_c$ superconductivity.  
In order to investigate the pairing mechanism, 
here we first construct an  
electronic model for LaFeAsO$_{1-x}$F$_x$ using 
maximally localised Wannier orbitals obtained from 
first principles calculation.  The minimal model 
turns out to involve all the five Fe $d$ orbitals.\cite{1stpaper} 
Hence the iron-based material is contrasted with the cuprate, 
which is a one-band, doped Mott insulator.  
We then apply the random-phase approximation (RPA) 
to solve the Eliashberg equation. 
We conclude that a nesting between multiple Fermi 
surface (pockets) results in a development of a peculiar 
spin fluctuation mode, which in turn 
realises an unconventional pairing, which is basically an extended
$s$-wave where the gap function changes sign across the nesting 
vector.\cite{1stpaper,Mazin}
The result is intriguing as a realisation of the
general idea that the way in which electron correlation 
effects appear is very sensitive to the underlying band structure 
and the shape of the Fermi surface.\cite{fermiology}

\section{Band structure}

\begin{figure}[h]
\begin{center}
\includegraphics[width=10.0cm,clip]{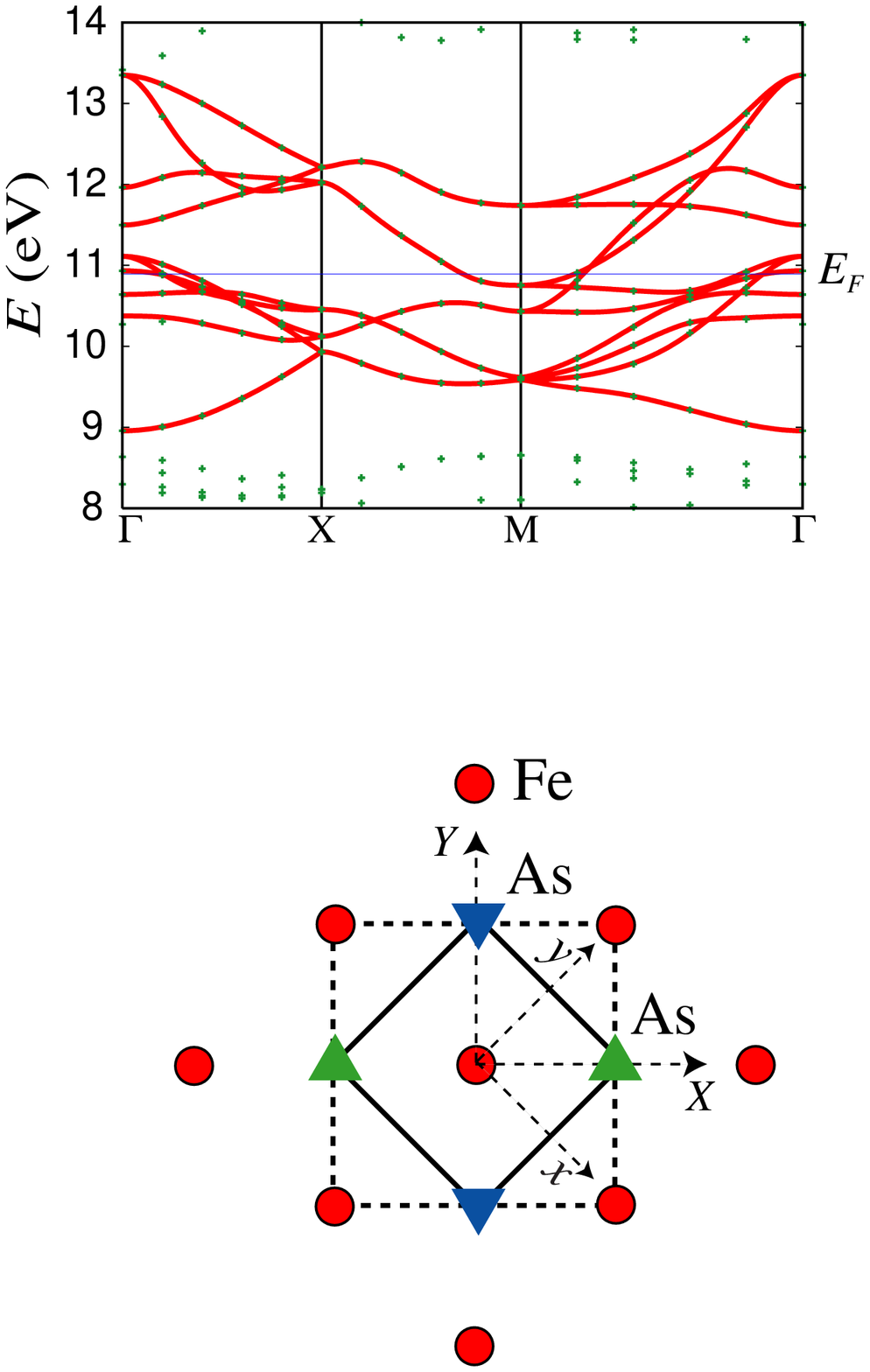}
\caption{Top panel: The band structure of in the original (folded) 
Brillouin zone, where 
the symbols represent the LDA result while curves the 
ten-band model derived with the maximally localised Wannier orbitals.  
Bottom panel: The original (dashed) and the 
reduced (solid) unit cells in real space with $\bullet$ (Fe), $\nabla$ (As 
below the Fe plane) and $\triangle$ (above). 
\label{fig1}}
\end{center}
\end{figure}

LaFeAsO has a layered structure, where Fe atoms form a 
square lattice in each layer, which is 
sandwiched by As atoms.\cite{2D3D} 
Due to the tetrahedral coordination of As atoms, 
there are two Fe atoms per unit cell.  
The experimentally determined lattice constants 
are $a=4.03$\AA \ and $c=8.74$\AA, with 
two internal coordinates $z_{\rm La}=0.142$ and $z_{\rm As}=0.651$.\cite{Cruz}
We have first obtained the band structure (Fig.\ref{fig1}) 
for these coordinates with the 
density-functional approximation with plane-wave basis\cite{pwscf},
which is then used to 
construct the maximally localised 
Wannier functions (MLWFs)\cite{MaxLoc}. 
These MLWFs, centered at the two Fe sites in the unit cell, 
have five orbital symmetries ($d_{3Z^2-R^2}$, 
$d_{XZ}$, $d_{YZ}$, $d_{X^2-Y^2}$, 
$d_{XY}$, where $X, Y, Z$ refer to those for this unit cell 
with two Fe sites as shown in the bottom panel of Fig.\ref{fig1}). 
The two Wannier orbitals in 
each unit cell are equivalent in that each Fe atom has the same 
local arrangement of other atoms.
We can then take a unit cell that 
contains only one orbital per symmetry by 
unfolding the Brillouin zone,\cite{comment}  
and we end up with an effective five-band model on a 
square lattice, where 
$x$ and $y$ axes are rotated by 
45 degrees from $X$-$Y$.  
We refer all the wave vectors in the unfolded Brillouin 
zone hereafter. 
We define the band filling $n$ as the number of electrons/number of sites
(e.g., $n=10$ for full filling). 
The doping level $x$ 
in LaFeAsO$_{1-x}$F$_x$ is related to the band filling as $n=6+x$.

\begin{figure}[h]
\begin{center}
\includegraphics[width=16.0cm,clip]{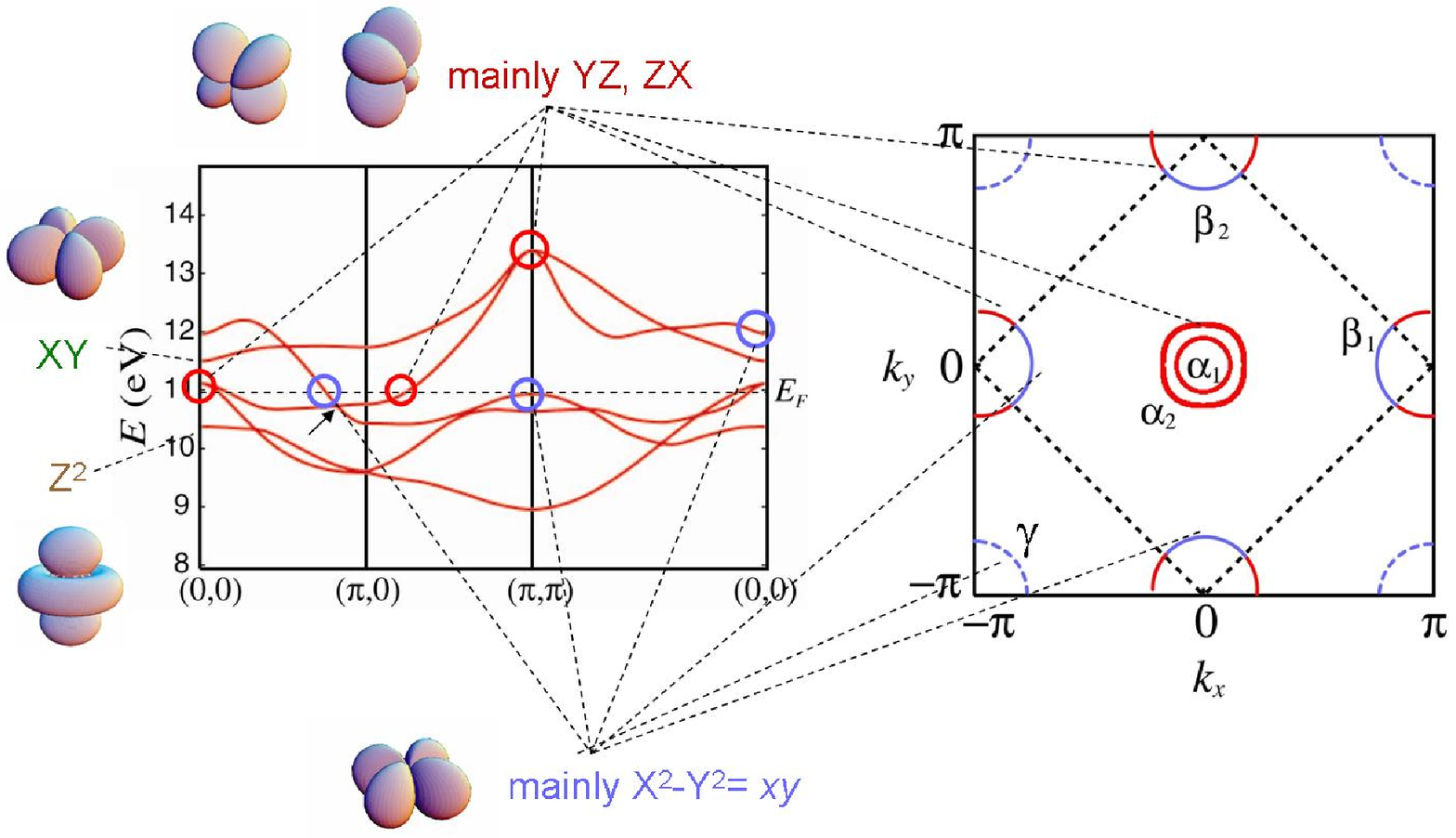}
\caption{Left: The band structure of the five-band model 
in the unfolded Brillouin zone. Orbital characters of the bands are 
also indicated. The short solid arrow denotes the position of the Dirac cone. 
Right: Fermi surface for $n=6.1$ in the unfolded Brillouin zone, 
where the original Bz is indicated by dashed lines. 
The blue (red) portions have strong 
$X^2-Y^2 (YZ, ZX)$ character. Dashed blue curves represent the 
portions where a $X^2-Y^2$ charactered 
band lies very close to the Fermi energy for $n=6.1$, which actually become a 
Fermi surface for smaller doping.
\label{fig2}}
\end{center}
\end{figure}

The five bands are heavily 
entangled as shown in Fig.\ref{fig2} (left panel)
reflecting strong hybridisation 
of the five $3d$ orbitals, which is physically due to the tetrahedral 
coordination of As atoms around Fe.
Hence we conclude that the minimal electronic model 
requires all the five bands. In Fig.\ref{fig2}(right), 
the Fermi surface for $n=6.1$ (corresponding to $x=0.1$) 
obtained by ignoring the inter-layer hoppings 
is shown in the two-dimensional unfolded Brillouin zone.

The Fermi surface consists of four pieces (pockets in 2D):   
two concentric hole pockets (denoted here as $\alpha_1$, $\alpha_2$) 
centered around $(k_x, k_y)=(0,0)$, two electron pockets 
around $(\pi,0)$ $(\beta_1)$ or $(0,\pi)$ $(\beta_2)$, respectively. 
$\alpha_i$ ($\beta_i$) corresponds to the 
Fermi surface around the $\Gamma$Z line (MA in the original Brillouin zone) 
in the first-principles band calculation.\cite{Singh}
Besides these pieces of the Fermi surface, there is a portion of the band 
near $(\pi,\pi)$ that is flat and 
touches the $E_F$ at $n=6.1$, so that 
the portion acts as a ``quasi Fermi surface $(\gamma)$'' around $(\pi,\pi)$, 
which has in fact an important contribution to the spin susceptibility. 
As for the orbital character, $\alpha$ and portions of $\beta$ near 
Brillouin zone edge have mainly $d_{XZ}$ and $d_{YZ}$ character, 
while the portions of 
$\beta$ away from the Brillouin zone edge and $\gamma$ have 
mainly $d_{X^2-Y^2}$ orbital character (Fig.\ref{fig3}, bottom panels).

An interesting feature in the band structure 
is the presence of Dirac cones, i.e., 
places where the upper and the lower bands make a conical contact.
\cite{Ishibashi,Fukuyama}
The ones closest to the Fermi level lies at positions where the 
$d_{X^2-Y^2}$ and the $d_{XZ}/d_{YZ}$ bands cross, just below the 
$\beta$ Fermi surface. 

\section{Many-body Hamiltonian and 5-band RPA}

We consider a two-dimensional model where the inter-layer hoppings 
are neglected.
For the many body part of the Hamiltonian, 
we consider the standard interaction terms that comprise 
the intra-orbital Coulomb $U$, the inter-orbital 
Coulomb $U'$, the Hund's coupling $J$ and the pair-hopping $J'$. 
The many body Hamiltonian then reads
\begin{eqnarray}
H &=& \sum_i\sum_\mu\sum_{\sigma}\varepsilon_\mu n_{i\mu\sigma}
 + 
\sum_{ij}\sum_{\mu\nu}\sum_{\sigma}t_{ij}^{\mu\nu}
c_{i\mu\sigma}^{\dagger} c_{j\nu\sigma}\nonumber\\
&+&\sum_i\left[U\sum_\mu n_{i\mu\uparrow} n_{i\mu\downarrow}
+U'\sum_{\mu > \nu}\sum_{\sigma,\sigma'} n_{i\mu\sigma} n_{i\mu\sigma'}
\right.\nonumber\\
&&\left.+J\sum_{\mu\neq\nu}\Vec{S}_{i\mu}\cdot\Vec{S}_{i\nu}
+J'\sum_{\mu\neq\nu} c_{i\mu\uparrow}^\dagger c_{i\mu\downarrow}^\dagger
c_{i\nu\downarrow}c_{i\nu\uparrow}
\right], 
\end{eqnarray}
where $i,j$ denote the sites and $\mu,\nu$ the orbitals, 
and $t_{ij}^{\mu\nu}$ is the transfer energy obtained in the 
previous section. The orbitals $d_{3Z^2-R^2}$, 
$d_{XZ}$, $d_{YZ}$, $d_{X^2-Y^2}$, 
$d_{XY}$ are labeled as $\nu=1,2,3,4,$ and 5, respectively.
As for the electron-electron interactions, there have been 
some theoretical studies that estimate the parameter values.
Some give $U\gg J$,\cite{Nakamura,Miyake} 
while others $U\sim J$.\cite{Anisimov}
We assume here that $U\gg J$, and take the values 
$U=1.2$, $U'=0.9$, $J=J'=0.15$ throughout the study. 
These values are smaller than the values 
obtained in ref.\cite{Nakamura,Miyake} because 
the self energy correction is not taken into account in the 
present calculation, so that small values of interaction parameters 
are necesssary to avoid magnetic ordering at high temperatures.

Having constructed the model, we move on to the RPA calculation, 
where the modification of the band structure due to 
the self-energy correction is not taken into account. 
Multiorbital RPA is described in e.g.
ref.\cite{Yada,Takimoto}. In the present case, 
Green's function $G_{lm}(k)$ $(k=(\Vec{k},i\omega_n))$
is a $5\times 5$ matrix. The irreducible susceptibility matrix 
\begin{equation} 
\chi^0_{l_1,l_2,l_3,l_4}(q) =\sum_q G_{l_1l_3}(k+q)G_{l_4l_2}(k)
\end{equation}
$(l_i = 1,...,5)$ has $25\times 25$ components, and 
the spin and the charge (orbital) susceptibility matrices are obtained 
from matrix equations, 
\begin{equation}
\chi_s(q)=\frac{\chi^0(q)}{1-S\chi^0(q)}
\end{equation}
\begin{equation}
\chi_c(q)=\frac{\chi^0(q)}{1+C\chi^0(q)}
\end{equation}
where 
\begin{equation}
S_{l_1l_2,l_3l_4},\;\; C_{l_1l_2,l_3l_4}
=\left\{\begin{array}{ccc}
U,& U &\;\; l_1=l_2=l_3=l_4\\ 
U',&-U'+J & \;\; l_1=l_3\neq l_2=l_4\\
J,&2U'-J,&\;\; l_1=l_2\neq l_3=l_4\\
J',&J'& \;\; l_1=l_4\neq l_2=l_3\end{array}  \right.
\end{equation}
We denote the largest eigenvalue of the spin 
susceptibility matrix for $i \omega_n=0$ 
as $\chi_s(\Vec{k})$. 
The Green's function and 
the effective singlet pairing interaction, 
\begin{equation}
V^s(q)=\frac{3}{2}S\chi^s(q)S-\frac{1}{2}C\chi^c(q)C+\frac{1}{2}(S+C),
\end{equation}
are plugged into the linearised Eliashberg equation, 
\begin{equation}
\lambda \phi_{l_1l_4}(k)=-\frac{T}{N}\sum_q
\sum_{l_2l_3l_5l_6}V_{l_1 l_2 l_3 l_4}(q)G_{l_2l_5}(k-q)\phi_{l_5l_6}(k-q)
G_{l_3l_6}(q-k).
\end{equation}
The $5\times 5$ matrix gap function $\phi_{lm}$ 
in the orbital representation 
along with the associated eigenvalue $\lambda$ are obtained by 
solving this equation. 
The gap function can be transformed into the band representation 
with a unitary transformation. 
The temperature is fixed at $T=0.02$eV throughout the study, and 
$32\times32$ $k$-point meshes and 1024 Matsubara frequencies are taken. 
We find that the spin fluctuations dominate in magnitude over orbital 
fluctuations as far as 
$U>U'$, so we can characterise the system with the spin susceptibility.

\section{Result: spin structure}

\begin{figure}
\begin{center}
\includegraphics[width=10.0cm,clip]{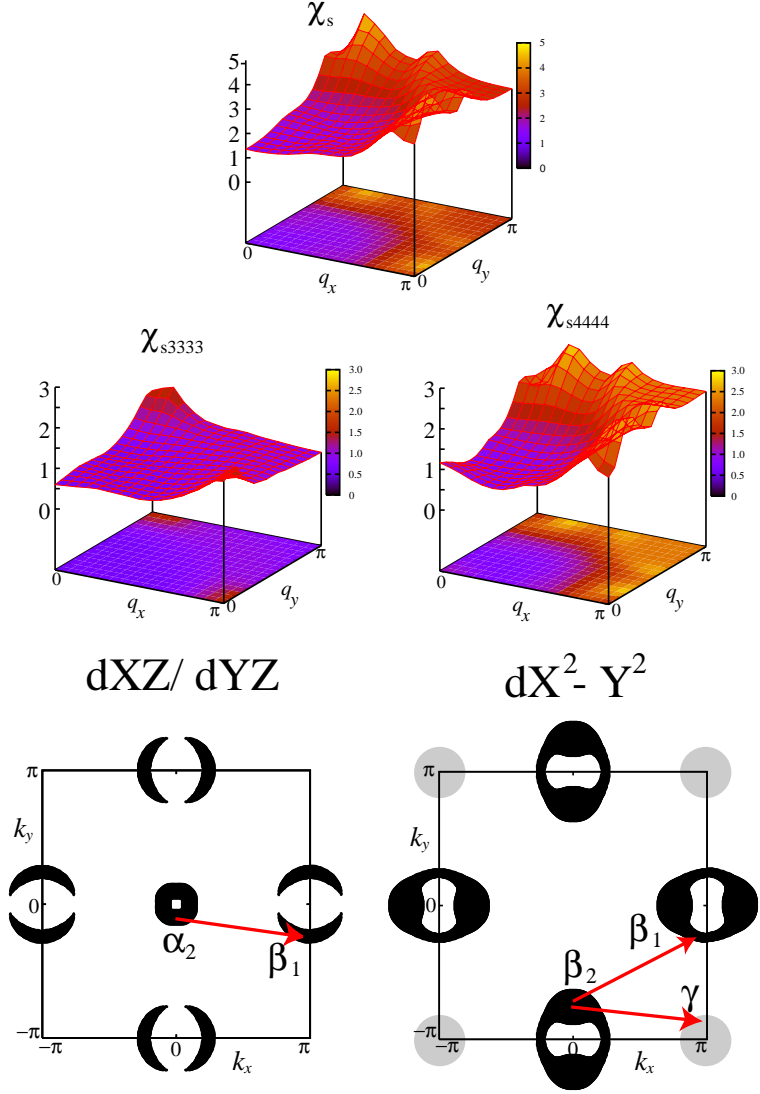}
\caption{
Top panel: Five-band RPA result for the spin susceptibility $\chi_s$   
for $U=1.2$, $U'=0.9$, $J=J'=0.15$, $n=6.1$ and $T=0.02$ (in eV).
Middle panels: Diagonal components, $\chi_{s3333}$(left) and 
$\chi_{s4444}$(right), of the 
spin susceptibility matrix in the orbital representation 
($3: YZ, 4: X^2-Y^2$).  
Bottom panel: Nesting is shown for the Fermi surface 
for orbitals $XZ, YZ$ (left)
and for $X^2-Y^2$ (right). Here the thickness of the Fermi surface 
represents the strength of the respective orbital character. The gray areas 
around the corners in the left panel indicates the 
$\gamma$ ``quasi'' Fermi surface. 
\label{fig3}}
\end{center}
\end{figure}

Let us first look at the key ingredient: the spin susceptibility 
in the top panel of Fig.\ref{fig3}. 
The susceptibility 
$\chi_s$  has peaks around 
$(k_x, k_y) = (\pi,0)$, $(0,\pi)$.  In addition we note that there is 
a ridge extending from $(\pi,\pi/2)$ to $(\pi/2,\pi)$ 
around $(\pi,\pi)$.
To explore the origin of these spin structures, 
we show $\chi_{s3333}$ and $\chi_{s4444}$ in the middle of 
Fig.\ref{fig3}, 
which represent the spin correlation 
within $d_{YZ}$ and $d_{X^2-Y^2}$ orbitals, respectively.
$\chi_{s3333}$ peaks solely around $(\pi,0)$ and $(0,\pi)$, 
which reflects the nesting between the 
$XZ,YZ$-charactered portions of 
$\alpha$ and $\beta$ Fermi pockets as shown in a bottom panel 
of Fig.\ref{fig3}.
On the other hand, $\chi_{s4444}$ has peaks around $(\pi,0),(0,\pi)$ 
and around $(\pi,\pi/2)/(\pi/2,\pi)$ as well. 
The former is due to the nesting 
between the $\gamma$ quasi Fermi surface and the $d_{X^2-Y^2}$ 
portion of the $\beta$ Fermi surface as was first pointed out in 
ref.\cite{Mazin},  while the latter originates 
from the nesting between the $d_{X^2-Y^2}$ 
portion of the $\beta_1$ and $\beta_2$ Fermi surfaces.\cite{1stpaper}
The $(\pi,0),(0,\pi)$ feature is 
consistent with the stripe (i.e., collinear) antiferromagnetic order for the 
undoped case, which was suggested by transport 
and optical reflectance,\cite{Dong} 
and further confirmed by neutron scattering experiments.\cite{Cruz}
The stabilization of such an antiferromagnetic ordering has also been 
pointed out in first principles calculations.\cite{Ishibashi,Dong,Mazin2}

\section{Result: superconductivity}
\begin{figure}
\begin{center}
\includegraphics[width=8.0cm,clip]{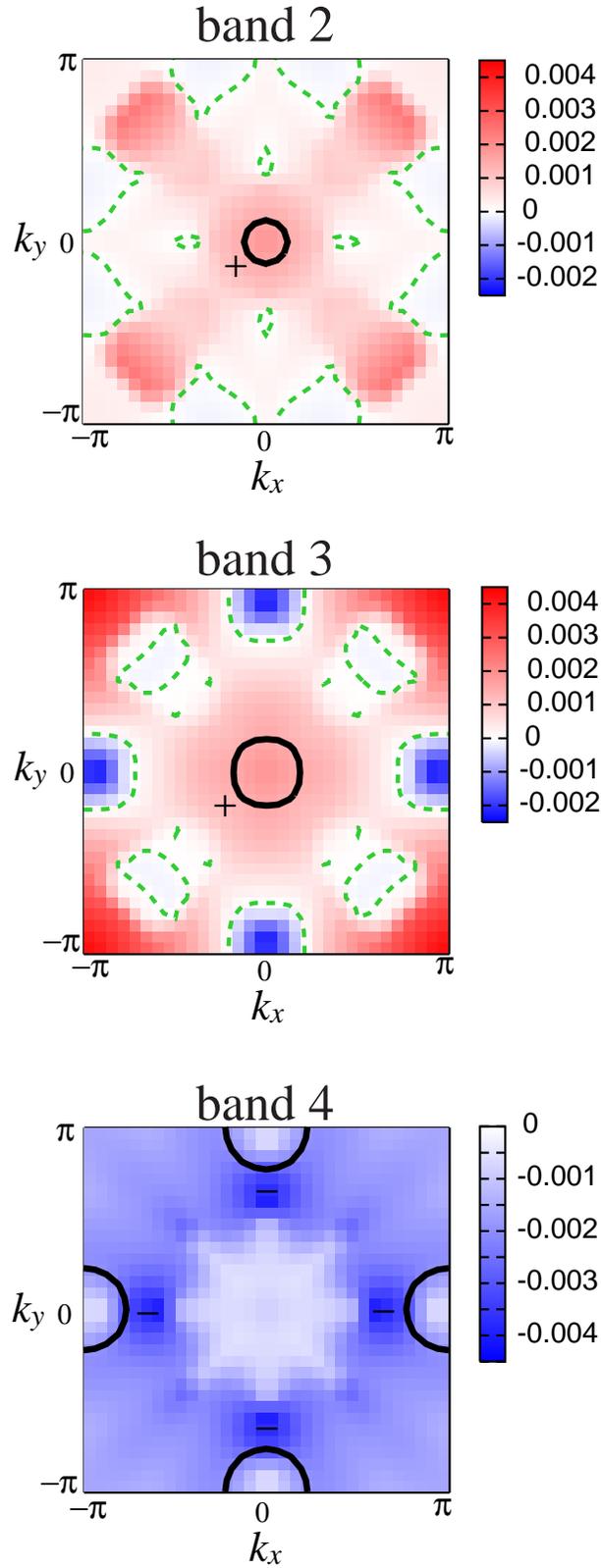}
\caption{Gap function in the band representation for 
band 2 (top) band 3 (middle) and band 4 for the same set of parameter values 
as in Fig.\ref{fig3}.
\label{fig4}}
\end{center}
\end{figure}

The presence of multiple set of nesting vectors revealed above 
provides an interesting case of the gap function in a 
spin-fluctuation mediated superconductivity, since 
multiple nestings can not only cooperate but also compete 
with each other.  
Namely, the $\alpha$-$\beta$ and $\gamma$-$\beta$ 
nestings tend to favour the sign reversing $s$-wave pairing, in 
which the gap changes sign between $\alpha$ and $\beta$ with 
a fixed sign (i.e., full gap) on each of the Fermi surface.\cite{Mazin}
On the other hand, $\beta_1-\beta_2$ nesting tends to change the 
sign of the gap between these two pockets, which can result in 
either $d$-wave pairing or an $s$-wave pairing with nodes on the 
$\beta$ Fermi surface\cite{Graser,errata}.
For the band structure of LaFeAsO (obtained by using the experimentally 
determined lattice structure), the $(\pi,0)$ spin fluctuation dominates, 
and the sign-reversing $s$-wave with no nodes on the Fermi pockets
dominates for the present set of parameter values.\cite{errata}
In Fig.\ref{fig4}, we show the gap function in the band representation 
for the third and the forth bands, which produce the $\alpha_2$ and $\beta$ 
Fermi surfaces, respectively.
A number of theoretical studies that adopt effective two band models,
\cite{Qi,Daghofer,Chubukov} 
the present five band model,\cite{Nomura,Ikeda,DHLee} 
or a 16 band $dp$ model\cite{Ono} 
have also found that this sign reversing $s$-wave is a 
good candidate of the gap function in this material.
The sign change in the $s$-wave gap 
is analogous to those in models studied by Bulut {\it et al.},\cite{Bulut} 
and also by the present author for the disconnected 
Fermi surfaces\cite{KA,KA2}.  
It is also reminiscent of the unconventional $s$-wave pairing 
mechanism for Na$_x$CoO$_2\cdot y$H$_2$O\cite{Takada} 
proposed by Kuroki et al.\cite{KKNCOO} 

We have to realise, however, that 
the gap can vary significantly along the $\beta$ Fermi surface, and 
also between different pieces of the Fermi surface 
due to the multiorbital character of the system.  
In the present case, the gap for the $d_{X^2-Y^2}$-charactered portion of the 
$\beta$ Fermi surface is about twice as large as that for 
the $XZ,YZ$-charactered portions, namely near the Brillouin 
zone edge of the $\beta$ Fermi surface and 
on the $\alpha_1$, $\alpha_2$ Fermi surface.  
However, we find, in further calculations, that this gap 
variance is not universal, and depends strongly on the electron density and 
also on the details of the band structure.  This is because 
the $\beta-\gamma$ nesting and thus 
the $(\pi,0)$ vs. $(\pi,\pi/2)$ spin fluctuation competition 
is sensitively affected by 
the position of the $d_{X^2-Y^2}$ portion of the 
band near $(\pi,\pi)$ with respect to the Fermi level. 
The details of this band filling and band structure dependences 
will be published elsewhere.

\section{Conclusion}
To summarise, we have constructed a five-band electronic 
model for LaFeAsO$_{1-x}$F$_x$, which we consider to be the 
minimum microscopic model for the iron-based superconductor. 
Applying a five-band RPA to this model,  
we have found that  spin-fluctuation modes around $(\pi,0),(0,\pi)$ 
develop due to 
the nesting between disconnected Fermi surfaces.
Based on the linearised Eliashberg equation, 
we have concluded that multiple spin fluctuation modes 
realises unconventional, extended $s$-wave pairing, where the 
gap changes sign across the nesting vectors. 

So the general picture obtained here is that 
the iron compound is a 
multi-band system having electron and hole pockets, as sharply 
opposed to the cuprate which is a 
one-band and nearly half-filled system with a simply connected 
Fermi surface.  
This poses a challenging future problem of 
elaborating respective pros and cons for the iron compound and the cuprate 
for superconductivity.

\ack
We wish to thank Ryotaro Arita, Seiichiro Onari, Hidetomo Usui, 
Yukio Tanaka, and Hiroshi Kontani for a collaboration 
in Ref.\cite{1stpaper} and for valuable discussions.  
Numerical calculations were performed at 
the Information Technology Center, University of Tokyo, 
and at the Supercomputer Center,
ISSP, University of Tokyo. 
This study has been supported by 
Grants-in-Aid for Scientific Research from  MEXT of Japan and from 
the Japan Society for the Promotion of Science.
%



\begin{thebibliography}{99}
\bibitem{comment}  While an ambiguity exists in unfolding the Brillouin zone 
[if the sign of the hoppings $t(\Delta x,\Delta y;\mu, \nu)$ with 
$\Delta x+\Delta y$=odd (i.e., hoppings between A and B sites in a bipartite 
lattice) is changed, a band structure is reflected with respect to  
$|k_x|+|k_y|=\pi$], 
this is just a unitary transformation, and either way the five-band structure 
gives the same ten bands in the folded Brillouin zone as well as the same 
RPA results.
\bibitem{errata} In our previous paper\cite{1stpaper} 
the gap function (diagonal element) 
of the 4th band had nodes intersecting the 
$\beta$ Fermi surface.  This was in fact due to a technical error in the 
final stage of the calculation, i.e., the unitary transformation from the 
orbital to the band representations.  
However, the main conclusions of Ref.\cite{1stpaper} 
do remain unaltered, since 
(i) the magnitude of the gap along the $\beta$ Fermi surface 
varies significantly as mentioned here, and 
(ii) the coexistence 
of $(\pi,\pi/2)$ and $(\pi,0)$ spin fluctuations determines the way 
in which nodes appear as asserted in \cite{1stpaper}.  
The nodes of the gap function 
do intersect for certain choices of the parameter values or a slight 
change in the band structure, as shown in ref.\cite{Graser}.
\end{thebibliography}
\end{document}